\newcommand{\diracslash}[1]{#1\llap{/\kern2pt}}

\newcommand{\be}{\begin{equation}}
\newcommand{\ee}{\end{equation}}
\newcommand{\bea}{\begin{eqnarray}}\index{\footnote{}}
\newcommand{\eea}{\end{eqnarray}}
\newcommand{\ba}[1]{\begin{array}{#1}}
\newcommand{\ea}{\end{array}}

\documentclass[epj,referee,onecolumn]{svjour} 
\usepackage{graphics}
\usepackage{amssymb}
\usepackage{amsfonts}
\begin{document}
\setlength{\topmargin}{0.2in}

\title{Electric field induced saturation effects in photoassociation of a pair of heteronuclear atoms}
\author{Debashree Chakraborty \inst{1} \and Bimalendu Deb \inst{1,2}}
\institute{Department of Materials Science, Indian Association for the Cultivation of Science,
 Jadavpur, Kolkata 700032, INDIA\and Raman Center for Atomic, Molecular and Optical Sciences, Indian Association for the Cultivation of Science,
Jadavpur, Kolkata 700032, INDIA}

\abstract
{We theoretically study saturation effects induced by an external static electric field on photoassociation (PA) of a heteronuclear atom pair. A static electric field influences scattering wave-function of two heteronuclear atoms as described in [D. Chakraborty, J. Hazra and B. Deb, J. Phys. B. {\bf 44} 095201 (2011)]. For certain values of electric field strengths there exist anisotropic resonances in ground state scattering leading to a large modification of scattering wave-function at short and intermediate separations where photoassociative Franck-Condon overlap is significant. Photoassociation rate as a function of collision energy shows a splitting near resonant electric field in the mili Kelvin energy regime. This splitting with a prominent dip occurs due to resonant enhancement of free-bound stimulated linewidth leading to saturation in free-bound transitions. We study electric field induced saturation effects on both one- and two-colour PA. Our results suggest that a static electric field may influence the formation of ground state polar molecules in two-colour PA.}

%pacs{34.50-s line and band widths}
\titlerunning{Electric field induced saturation effects in photoassociation }
\authorrunning{D. Chakraborty \it et al. }

\maketitle
\section{Introduction}
\label{intro}
Cold molecules below millikelvin temperature regime allow for a wealth of interesting studies. This is especially true for heteronuclear molecules due to their potential applications in numerous areas such as quantum computation \cite{rabl}, precision measurements of fundamental constants \cite{zelevinsky} and quantum phases of dipolar gases \cite{micheli,pupillo,dieckmann}. Several experimental groups have reported photoassociative formation of ultracold alkali-metal polar molecules, such as NaCs \cite{bigelow}, KRb \cite{marcassa,stwalley,wang,k.k.ni}, RbCs \cite{kerman,sage}, LiCs \cite{kraft,deiglmayr} and LiK \cite{voigt} in their electronic ground states.
Because of the existence of permanent dipole moment, polar molecules have significant long-range electric dipole-dipole interaction. This makes cold polar molecules an interesting system for exploring many-body physics with long-range interaction.
Not only polar molecules, but mixtures of two heteronuclear species of cold atoms are also of interest for several reasons.
First, in order to form polar molecules from cold atoms, 50:50 mixtures of two heteronuclear atomic gases serve as a starting point.
Second, heteronuclear mixtures of alkali metal atoms are amenable to magnetic field induced Feshbach resonances \cite{wille,tiecke,levinsen} and so offer a new scope for studying exotic physics of strongly interacting gases.
Third, since two heteronuclear cold atoms in quasimolecular collision complex posses a spatially dependent permanent dipole moment, the collisional properties of a pair of heteronuclear atoms can be manipulated by an external static electric field.
It is shown that, unlike in homonuclear atoms, a static electric field can significantly affect the interaction of a heteronuclear atom-pair \cite{krems,debashree} due to the existence of permanent dipole moment of collision complex.
The effect of a static electric field on the formation of ground state polar molecules via one-photon stimulated emission from ground continuum has been theoretically studied \cite{rosario}.

In a previous paper \cite{debashree}, we have studied the effects of a static electric field on photoassociation (PA) of $^7$Li + $^{133}$Cs into LiCs polar molecules in excited electronic state. A static electric field can couple states of different angular momenta (partial waves) of a heteronuclear pair of ground state cold atoms leading to anisotropic scattering resonances. The permanent dipole moment of $^7$Li + $^{133}$Cs collision complex has strong spatial dependence at short separations and diminishes at large separations. Therefore, an external static electric field predominantly affects scattering wavefunction of two heteronuclear atoms at short separations. In particular, the modification of scattering wavefunction is most prominent near electric fields at which anisotropic resonances appear. PA rate near such electric field-induced resonances is found to be enhanced. Furthermore, as a result of anisotropy of the ground scattering state, higher rotational levels of excited molecules can be populated by PA near the anisotropic resonances \cite{debashree}.
Thus an external static electric field can serve as a tool for manipulation of various physical and chemical processes of dipolar systems. 
%===========================================================================================================
% Figure- 1
\begin{figure}
\resizebox{1.0\columnwidth}{!}{%
  \includegraphics{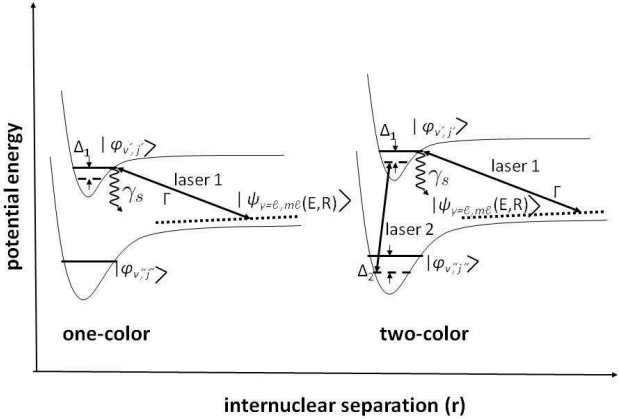}
}
%\vspace*{1cm} 
\caption{Schematic diagram of one- and two-colour photoassociation process. One-colour photoassociation process creates a free-bound-bound transition pathway between a colliding pair of atoms and an excited molecule. Here laser-1 is tuned to a excited bound molecular state $\mid \phi_{v', j'}\rangle$ from the continuum state $\mid \psi_{\gamma= l,m_l}\rangle$, $\Gamma$ is the free-bound stimulated linewidth and $\gamma_{s}$ corresponds to the spontaneous decay from state $\mid \phi_{v', j'}\rangle$. In case of two-colour PA, laser-2 couples the excited bound $\mid \phi_{v' , j'}\rangle$ and the ground bound state $\mid \phi_{v'' , j''}\rangle$. $\Delta_1$ and $\Delta_2$ are the detunings.}
\label{fig:1}
\end{figure}
%============================================================================================================  

Our main interest here is to investigate the saturation effects due to a static electric field on one and two-colour PA as schematically shown in figure 1. 
In one-colour PA, a laser of angular frequency $\omega_1$ is used to photoassociate a pair of colliding atoms into an excited molecular level with energy E$_{v', j'}$ where $v'$ and $j'$ are the vibrational and rotational quantum numbers, respectively. In two-colour PA, an additional laser of angular frequency $\omega_2$, can drive the dimer from bound level $(v', j')$ to a second bound level $(v'', j'')$ with energy $E_{v'', j''}$ in the ground electronic configuration.
PA spectrum is described as the rate of loss of atoms from the excited molecular level populated by PA process. Saturation occurs when free-bound stimulated linewidth exceeds spontaneous linewidth.
At saturation, the excited molecule has finite probability of going back to the initial two-atom continuum by stimulated emission. As a result, a free-bound coherence can be established at saturation. Therefore, PA rate (loss of atoms) is expected to exhibit a dip at saturation.

Here we show that, even with weak PA lasers, saturation can be achieved by tuning electric field near anisotropic resonances. This occurs due to large enhancement of free-bound Franck-Condon (FC) overlap near anisotropic resonances. 
Since free-bound stimulated linewidth is proportional to the square of FC integral, a large enhancement of this integral due to anisotropic resonances leads to saturation. 
Due to saturation induced by resonant electric field, a splitting is observed in one-colour PA spectra. In case of two-colour PA, when the bound-bound transition is tuned near one of the peaks of one-colour PA at saturation, a furthur splitting occurs resulting in a three peak PA spectrum.
Our results show that, in the absence of electric field, one- and two-colour PA probability is negligibly small in comparison to that in the presence of electric field induced anisotropic resonances.
In case of two-colour PA, state selectivity may be possible due to electric field.

This paper is organized as follows. First, we give a brief discussion of the effects of an external static electric field on the heteronuclear collision. We discuss some salient features of PA of a heteronuclear atom pair in presence of a static electric field in section 2. In section 3, we present and discuss numerical results. Finally, we summarize and conclude in section 4.
\section{PA in the presence of electric field}
\label{sec:1}
Before we discuss saturation effects due to a static electric field, let us first discuss how a static electric field affects ground state scattering properties of two heteronuclear atoms. 
In the absence of an electric field, different partial-wave states of a heteronuclear collision complex are uncoupled, and therefore the scattering is isotropic governed mainly by $s$-wave at ultralow kinetic energies. The interaction of the permanent dipole moment of a heteronuclear collision pair with a static electric field can be expressed as 
\bea
\hat V_{\cal E}(R) = -\cal E \rm cos\theta\sum_S\sum_{M_S}\mid S M_S\rangle d_S(R)\langle S M_S\mid\eea
where $\cal E$ is the electric field magnitude and $d_S$(R) the dipole moment function of the complex in the different electronic spin states S. $M_S$ is the projection of S on the Z axis.
The analytical expression of $d_S$(R) is given in \cite{krems}. The potential $\hat V_{\cal E}$(R) induces coupling between different partial waves $\ell$ such that even and odd $\ell$ are coupled separately to each other but not to themselves because of electric dipole selection rules. Due to this electric field induced anisotropic coupling, several scattering resonance structures appear \cite{debashree}. Figure 3 of \cite{debashree} shows that for $^7$Li + $^{133}$Cs collision we get the first resonance peak near electric field $\cal E$ = 1298 kV/cm and the second one near $\cal E$ = 1650 kV/cm. As mentioned earlier, the permanent dipole moment function of a heteronuclear collision complex has strong spatial dependence at relatively short separations. Hence an external static electric field interacting with the permanent dipole moment of a heteronuclear collision complex strongly modifies scattering wavefunction at short separations, especially near electric fields at which resonances occur. 
This electric field induced modification of two-body scattering can be contrasted with magnetic field induced alteration of atom-atom interaction. In presence of magnetic Feshbach resonances \cite{wille,tiecke,levinsen}, scattering wavefunction gets modified mainly at large separations through Zeeman effect but electric field can modify the wavefunction at short separations.
%===========================================================================================================
% Figure- 2
\begin{figure}
\resizebox{1.0\columnwidth}{!}{%
  \includegraphics{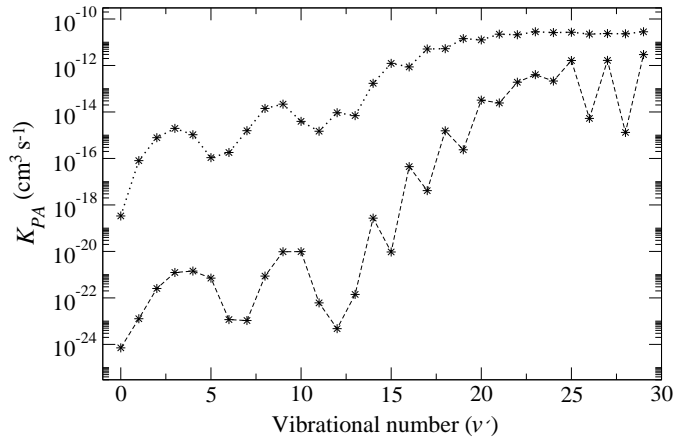}
}
\caption{The photoassociation rate $\textit{K}_\textit{PA}$ (in unit of cm$^3$s$^{-1}$) for the one-colour process is plotted as a function of vibrational number ($v'$) of the excited $B^1\Pi$ state at resonant electric field $\cal E$ = 1298 kV/cm (dotted line) and at zero electric field (dashed line). }
\label{fig:2}
\end{figure}
%===========================================================================================================

Next, we discuss how PA of two heteronuclear atoms is influenced by the electric field induced modification of ground state scattering wavefunction. 
For quantitative analysis, we have chosen PA transitions to the excited state B$^1\Pi$ of LiCs molecule. Because, PA transitions to this state occurs at relatively short separations and hence external electric field is expected to affect significantly PA spectrum of this state.

PA rate coefficient is given by
\bea K_{PA} = \left\langle\frac{\pi v_{rel}}{k^2}\sum_{\ell=0}^{\infty}(2\ell +1)\mid S_{PA}^{n}\mid^2\right\rangle\eea 
Where $S_{PA}^{n}$ \cite{napolitano,julienne} represents S-matrix element for one- ($n$ = 1) or two-colour ($n$ = 2) PA transition and $\langle....\rangle$ implies an averaging over thermal velocity distribution.
For sufficiently low laser intensities, the one-photon PA probability is given by
%===========================================================================================================
% Figure- 3
\begin{figure}
\resizebox{1.0\columnwidth}{!}{%
  \includegraphics{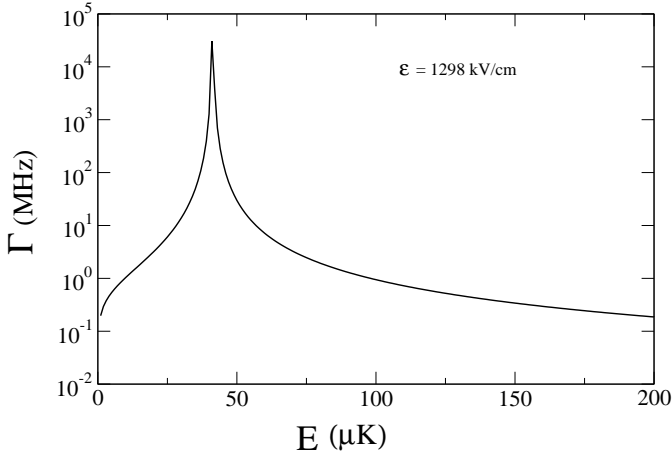}
}
\caption{The stimulated linewidth $\Gamma$ (in MHz) is plotted as a function of energy E (in $\mu$K) at the first resonant electric field $\cal E$ = 1298kV/cm at I$_1$ = 1mW/cm$^2$. A peak arises at $\Gamma/\gamma_s \sim$ 10$^3$.}
\label{fig:3}
\end{figure}
%============================================================================================================
\bea \mid S_{PA}^{1}(E,\ell,\omega_1,I_1)\mid^2 = \frac{\hbar^2\gamma_s\Gamma(E,\ell,I_1)}{[(E-\Delta_1)^2+\hbar^2((\Gamma+\gamma_s)/2)^{2}]}\eea 
where $\gamma_s$ represents the natural linewidth of the excited state and the detuning of laser-1 is given by $\Delta_1 = E_{v', j'}-\hbar\omega_1$. 
The partial stimulated linewidth $\Gamma(E,\ell,I_1)$ is given by 
\bea \Gamma(E,\ell,I_1) = \frac{\pi I_1}{\epsilon_{0}c}\mid\langle \phi_{v',j'}\mid D_t(R)\mid \psi_{\gamma = l,m_l}(E,R)\rangle\mid^2\eea
where $D_t(R)$ is the transition dipole moment, $\epsilon_{0}$ and c are the vaccum permitivity and speed of light, respectively. $\phi_{v',j'}$ represents the excited bound state and $\psi_{\gamma = l,m_l}$ is the partial wave scattering wavefunction in the presence of a static electric field \cite{debashree}. The two-colour stimulated Raman PA probability is given by 
%===========================================================================================================
% Figure- 4
\begin{figure}
\resizebox{1.0\columnwidth}{!}{%
  \includegraphics{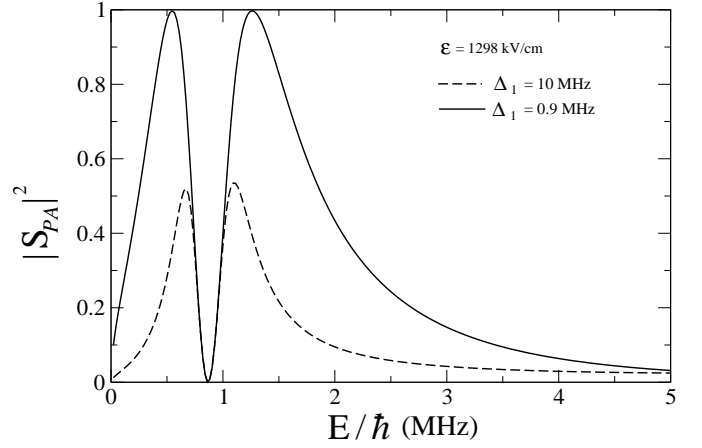}
}
\caption{The one-colour PA probability $\mid S_\textit{PA}\mid^2$ is plotted as a function of collisional energy E/$\hbar$ (in MHz) at the resonant electric field $\cal E$ = 1298kV/cm for $\Delta_1$ = 10 MHz (dashed line) and $\Delta_1$ = 0.9 MHz (solid line). A splitting is observed in the PA spectra due to saturation effect.}
\label{fig:4}
\end{figure}
%============================================================================================================
\bea 
&\mid S_{PA}^{2}(E,\ell,\omega_1,\omega_2,I_1,I_2)\mid^2& \nonumber\\ &=
\frac{\hbar^2(E-\Delta_2)^2\gamma_s\Gamma}{[(E-\Delta_+)(E-\Delta_-)]^2 + \hbar^2((\Gamma+\gamma_s)/2)^{2}(E-\Delta_2)^2}&.
\eea 
The second laser thus splits a single PA resonance into a pair of peaks located near the energies 
\bea
 \Delta_{\pm} = \frac{1}{2}(\Delta_1+\Delta_2)\pm\frac{1}{2}\sqrt{(\Delta_1-\Delta_2)^2 + 4\hbar^{2}\Omega_{12}^2}.
\eea 
Laser-2 is detuned by $\Delta_2 = E_{v'', j''}-\hbar(\omega_1-\omega_2)$ from $E_{v'', j''}$.
Here $\Omega_{12}$ = $\Omega_{12}^{0} C_{j'j''}$ where
\bea\Omega_{12}^{0} = \frac{1}{\hbar}\left(\frac{I_2}{4\pi\epsilon_0c}\right)^{1/2}\mid\langle\phi_{v'',j''}\mid D_t(R)\mid\phi_{v',j'}\rangle\mid\eea
and $\mid\phi_{v'',j''}\rangle$ is the ground bound state wavefunction. $C_{j'j''}$ represents the angular part which is given by 
\bea
C_{j'j''}=\sqrt{\frac{(2j'+1)}{(2j''+1)}}\langle j'm_{j'},1m\mid j''m_{j''}\rangle\nonumber\\\times\langle j'\Omega',1q\mid j''\Omega''\rangle\eea
where $\langle..\mid..\rangle$ represents the Clebsch-Gordan coefficient. $\Omega'$ and $\Omega''$ are the projections of the total electronic angular momentum of the excited and ground state, respectively, on the molecular axis whereas $m_j'$ and $m_j''$ are the projections onto the space-fixed frame. The quantum numbers $q$ and $m$ correspond to the projection of the dipole operator onto the body-fixed and space-fixed frames, respectively.
The molecular Rabi coupling is proportional to the bound-bound Franck-Condon(FC) factor.  
The equations (3), (4) and (5) indicate that PA rate of equation (2) is mainly affected by the electric field through the free-bound stimulated linewidth.
%===========================================================================================================
% Figure- 5
\begin{figure}
\resizebox{1.0\columnwidth}{!}{%
  \includegraphics{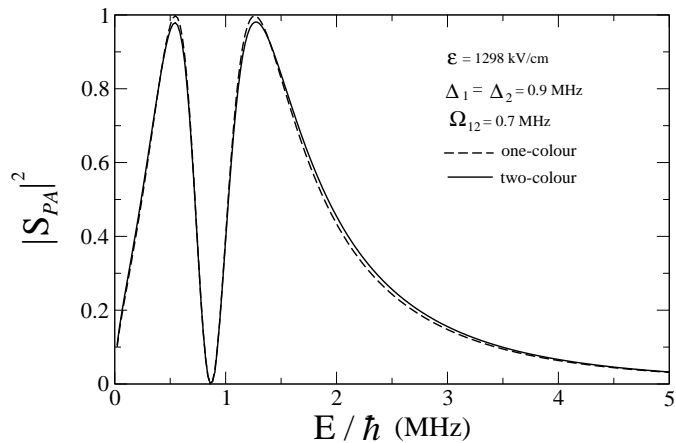}
}
\caption{$\mid S_\textit{PA}\mid^2$ is plotted as a function of collisional energy E/$\hbar$ (in MHz) at the resonant electric field $\cal E$ = 1298kV/cm for one-colour (dashed line) as well as two-colour (solid line) PA process. The detunings are set at $\Delta_1$ =  $\Delta_2$ = 0.9 MHz and $\Omega_{12}$ = 0.7 MHz.}
\label{fig:5}
\end{figure}
%============================================================================================================
\section{Results and discussions}
\label{sec:2}
Here we investigate saturation effects due to electric field in both one- and two-colour PA of Li + Cs.
In figure 2 the one-colour PA rate is plotted as a function of vibrational number ($v'$) at both zero and the resonant electric field ($\cal E$ = 1298kV/cm). We have joined the discrete values (as indicated by stars) by dashed line (zero electric field) and dotted line (resonant electric field) to guide readers' eyes. One-colour PA rate gets strongly enhanced by electric field induced anisotropic resonances. This figure further shows that the electric field effects are more prominent in case of lower vibrational states. 
We have chosen the excited state $v'= 26, j'= 1$ of the $B^1\Pi$ potential since it has a large Franck-Condon overlap with the ground continuum at $\cal E$ = 1298kV/cm. For the chosen excited state $v'=26$ we get the maximum overlap for ground vibrational state $v''=28$. The selection rule for the rotational transition is $\Delta J=0, \pm 1$ with the restriction $j=0\nrightarrow j=0$.  
In figure 3 we have plotted the stimulated linewidth $\Gamma$ (in MHz) as a function of energy E (in $\mu$K) at the resonant electric field $\cal E$ = 1298kV/cm. $\Gamma$ exceeds the spontaneous linewidth $\gamma$ by 3 orders of magnitude at collision energy E $\approx$ 41$\mu$K.
This large enhancement of $\Gamma$ is due to electric field induced resonant enhancement of FC integral.
The one-colour PA rate as given in equation (2) basically describes the loss of atoms due to spontaneous emission \cite{napolitano,julienne,bohn} from the excited molecular state. In case of $\Gamma\gg\gamma_s$, saturation occurs in free-bound transition.
In such a situation the excited molecule formed by PA can be coherently disintegrated by stimulated emission recycling the two atoms back to the same ground continuum.
This is worthwhile to stress that this strong coupling is brought in by electric field even at low laser intensity.
The loss of atoms due to spontaneous emission from excited bound state $\mid \phi_{v', j'}\rangle$ is expected to be thwarted or suppressed due to saturation effects.
%===========================================================================================================
% Figure- 6
\begin{figure}
\resizebox{1.0\columnwidth}{!}{%
  \includegraphics{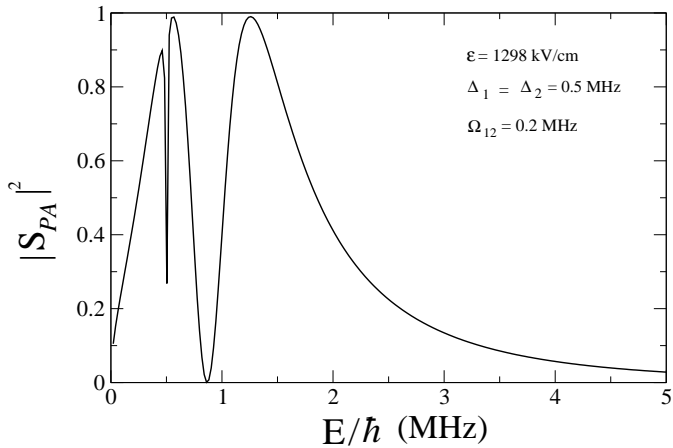}
}
\caption{$\mid S_\textit{PA}\mid^2$ is plotted as a function of collisional energy E/$\hbar$ (in MHz) for two-colour PA for $\cal E$ = 1298kV/cm and $\Omega_{12}$ = 0.2 MHz. The detunings are set at $\Delta_1$ =  $\Delta_2$ = 0.5 MHz which corresponds to the energy of the first peak of figure 4.}
\label{fig:6}
\end{figure}
%============================================================================================================

For numerical illustration, we first solve ground state scattering wavefunction. Since in the presence of a static electric field, differnt partial wave states are coupled we need to compute anisotropic scattering wavefunction.
To solve the coupled differential equations, Numerov-Cooley algorithm-based multichannel scattering technique was used. The bound state wavefunctions are solved using renormalized Numerov method. 
PA probability as a function of collision energy shows a splitting at resonant electric field in the micro kelvin energy regime as displayed in figure 4 and 5. The dip of the splitting appears at 41$\mu$K at which $\Gamma$ is maximized.
This dip results from the lowering of PA rate due to saturation.
At saturation, atoms can be recycled back to the continuum thwarting PA rate.
%===========================================================================================================
% Figure- 7
\begin{figure}
\resizebox{1.0\columnwidth}{!}{%
  \includegraphics{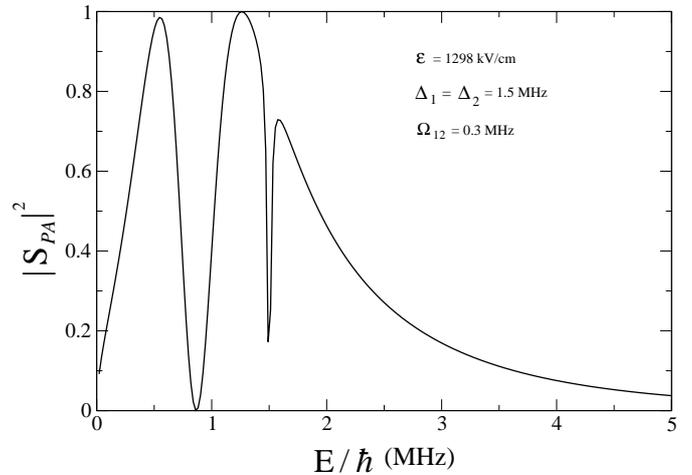}
}
\caption{$\mid S_\textit{PA}\mid^2$ is plotted as a function of collisional energy E/$\hbar$ (in MHz) for two-colour PA for $\cal E$ = 1298kV/cm and $\Omega_{12}$ = 0.3 MHz. The detunings are set at $\Delta_1$ =  $\Delta_2$ = 1.5 MHz which corresponds to the energy of the second peak of figure 4.}
\label{fig:7}
\end{figure}
%============================================================================================================
In figure 4 we have plotted the one-colour PA probability as a function of collisional energy E (in MHz) for $\Delta_1$ = 10 MHz (dashed line) and 0.9 MHz (solid line) at the resonant electric field at laser intensity I$_1$ = 1mW/cm$^2$. It is observed that the peak position as well as peak height changes with changing detuning but the position of the dip remains the same.
With the increase of the detuning of laser-1, height of the two peaks decreases symmetrically. Figure 5 shows two-colour Raman PA for $\Delta_1$ =  $\Delta_2$ = 0.9 MHz which corresponds the energy at the dip. Hardly any change is observed in this two-colour (solid line) PA spectra in comparison to one-colour (dashed line) PA at $\Delta_1$ = 0.9 MHz.      
Then we have set the detunings at the energies of the two-peaks and found three peak structures in the two-colour PA spectra as shown in figure 6 and figure 7 . In figure 6, the detunings are $\Delta_1$ =  $\Delta_2$ = 0.5 MHz and $\Omega_{12}$ = 0.2 MHz, while in figure 7 $\Delta_1$ =  $\Delta_2$ = 1.5 MHz and $\Omega_{12}$ = 0.3 MHz. These parameters correspond to the energy of the first and second peak, respectively. 
It is interesting to note that the extra dip appearing in two-colour PA has a shape resembling that of electromagnetically induced transparency (EIT).
However, unlike in an ideal EIT situation, this dip here does not go to zero. The sharp feature of the second dip is an indication that with the use of static electric field induced anisotropic resonances under appropriate detuning conditions at ultralow temperatures,
ground state polar molecules may be formed in some specific selective rovibrational states by two-colour PA. This is not possible in the absence of electric field for the case Li + Cs PA since PA probability in the absence of electric field is negligibly small, particularly for low lying vibrational levels, as revealed in figure 2.
\section{Conclusion}
\label{sec:3} 
In conclusion, we have studied saturation effects in one- and two-colour PA of two heteronuclear atoms induced by a static electric field. 
Forming ground state polar molecules in specific ro-vibrational states by two-colour coherent Raman PA is an important issue in cold atom science.
Although, it is extremely difficult to attain state selectivity in PA of a thermal gas \cite{pillet}, nevertheless a static electric field induced ground state scattering resonances may be useful in bringing state selectivity through enhanced free-bound Franck-Condon overlap integral. In the specific case of Li + Cs system, we have demonstrated that near electric field induced resonances, a three peak spectrum appears in two-colour PA.
Among the two dips of the three-peak spectrum, one is due to saturation effect and the other is due to quantum interference of two transition pathways of two-colour PA. 
Saturation effects and state selectivity due to electric field induced resonances can be more rigorously investigated by calculating dressed state of two bound states and a continuum in two-colour PA.
Bound-continuum-bound dressed state due to two lasers has been studied recently in a different context \cite{bdeb}.
It would be interesting to explore the possibility of a dark resonance in two-colour Raman PA due to static electric field effects.

\section{Acknowledgment}
\label{sec:4}
DC is grateful to CSIR, Government of India, for a support.   
%\section*{References}

\end{document}